# Lorentz Electrodynamics, Matter/Antimatter Cosmology and Astroparticle Physics

Anatoly Blanovsky
Teacher technology Center
7850 Melrose Ave., Los Angeles, CA 90046
Tel. 323-653-4085, FAX 323-658-6077

## Abstract
It is shown that the noncommutative Lorentz metric satisfies so-called nonpropagating waves. The electromagnetic and gravitational forces are obtained as a description of these wave motions in a certain dispersive medium. This approach leads to the natural introduction of the field values (group velocity and intensity of de Broglie waves) into Maxwell's equations and antimatter into Newtonian mechanics. This paper illustrates the important theoretical motivations of using radio astronomy technique for the Sagnac effect and cosmic ray study. One of the early predictions of the wave theory was existence of antiparticles (modes with negative group velocity). They are attracted by ordinary matter, but repulse ordinary matter and each other gravitationally. The interaction between the two types of matter could lead to expansion of the universe and maintain the energy reservoir for cosmic sources of intense radiation.

## Introduction

It is presently believed that physical experiments may detect acceleration, for example a rotation (Sagnac experiment, Foucault pendulum, etc.), but no physical experiments can detect a uniform translation. The anisotropy of the cosmic microwave background contradicts to the results of the traditional relativistic experiments, and we cannot see any evidence of earth orbital rotation from measuring the Sagnac effect on open paths.

There is also a difficulty with the mathematical formulation modern theory of electromagnetic wave propagation that is based on the postulate about constancy of the light velocity. In continuum mechanics, the Lorentz transformation is a mathematical tool that preserves a wave equation in a moving reference system instead of time invariance in Galilean transformation. As it is true for all wave motions, in all applications of the Lorentz transformation anisotropy values of characteristic wave velocity are used.

The most important difference between the Lorentz and Galilean transformations -noncommutative character of the Lorentz group is connected with wave properties of the observed phenomena. Historically, a long time before quantum mechanics, Fogt introduced Lorentz transformation in the elastic light theory. Then Lorentz came with a wave concept of relativity. He supposed that some disturbances, like waves, could be transmitted with traveling particles through a certain medium-ether without moving it.

E-mail ablanovs@lausd.k12.ca.u

## Lorentz Group and Dispersive Waves

In classical field theory, the Lorentz transformation could be derived as local relations-differential equations relating the coordinates of two reference systems, K and K', in relative motion [1]. We have

$$dt' = \frac{\partial t'}{\partial t}dt + (\vec{V} \cdot grad\, t')dt = \frac{1}{\sqrt{1 - V^2/c^2}}(dt - \frac{d\vec{r}\vec{V}}{c^2}), \quad (1)$$

$$d\vec{r}' = \frac{1}{\sqrt{1 - V^2/c^2}}(\frac{d\vec{r}\vec{V}}{V^2}\vec{V} - \vec{V}dt) + d\vec{r} - \frac{d\vec{r}\vec{V}}{V^2}\vec{V}, \quad (2)$$

Here $\vec{V}(\vec{r},t)$ is the velocity at a point fixed in reference system K', and c is a real constant that has units of velocity.

From the demand that dt' is the total derivative, we have

$$\frac{\partial \vec{V}}{\partial t} = -(\vec{V} \cdot grad)\vec{V} = \lambda(\vec{r},t)\vec{V}, \quad (3)$$

where $\lambda$ is an arbitrary function. Since the total derivative of the vector of velocity is equal zero, the considered frames are inertial.

In a reference system moving in the x direction at velocity v, it is convenient to express the Lorentz transformation in the two-dimensional coordinates $\xi$=x-ct and $\eta$=x+ct

$$\xi' = \sqrt{\frac{1-\beta}{1+\beta}}\xi = e^{\theta} \text{ and } \eta' = \sqrt{\frac{1+\beta}{1-\beta}}\eta = e^{-\theta}, \quad (4)$$

where $\beta$=V/c is a dimensionless constant and $\theta$=tanh$^{-1}\beta$. We have that the Lorentz transformation expands one coordinate while contracting the other to preserves their product or a wave equation in a moving reference system.

We have from the equation (1), the function t' satisfies the eikonal equation

$$\left(\frac{\partial t'}{\partial t}\right)^2 - \frac{c^4}{V^2}\left[(\frac{\partial t'}{\partial x_1})^2 + (\frac{\partial t'}{\partial x_2})^2 + (\frac{\partial t'}{\partial x_3})^2\right] = 0, \quad (5)$$

or Hamilton-Jacobi equation of relativistic mechanics in dimensionless form

$$c^2\left[(\frac{\partial t'}{\partial x_1})^2 + (\frac{\partial t'}{\partial x_2})^2 + (\frac{\partial t'}{\partial x_3})^2\right] - (\frac{\partial t'}{\partial t})^2 + 1 = 0. \quad (6)$$

In general case, the dispersion relation is determined by particular equations of the problem. For linear equation it is simply based on correspondence $\partial/\partial t \leftrightarrow -i\omega$, $\partial/\partial x \leftrightarrow k$, where k is a wave vector and $\omega$ is a frequency. Derivatives of the function t' are connected by dispersion relations $\omega^2=c^2k^2+\omega_c^2$, where $\omega_c$ is a constant. These waves are known in classical continuum mechanics as the nonpropagation waves.

Major characteristics of these waves are: 1) They are described by the Klein-Gordon equations. 2) The group and phase velocity are related by vu=c². 3) The group velocity is always less than characteristic velocity c. 4) The group velocity approaches zero for frequency below $\omega_c$, which is called the cutoff frequency. 5) The wave system becomes nondispersive in the limit k→∞ with v=u=c [2].

These waves also occur in vacuum. In this case $\omega_c = mc^2/\hbar$, where $\hbar$ is Plank constant, c is the velocity of light and m is mass of an associated particle. Multiply the derivatives of time by $\omega_c$, we have for a wave associated with a free particle that the elementary phase $d\theta = -(\frac{mc^2}{\hbar})\sqrt{1 - v^2/c^2}\, dt$, the wave vector $\vec{k} = (\frac{m\vec{v}}{\hbar})/\sqrt{1 - v^2/c^2}$ and the frequency $\omega = (\frac{mc^2}{\hbar})/\sqrt{1 - v^2/c^2}$.

### Propagation of Wave Disturbances and Path Integrals

The linear combination of the waves can be made to satisfy any boundary and initial conditions. In general, the Fourier integrals give exact solutions, but their content is difficult to see. To understand the main features of nonpropagation waves we consider the asymptotic behavior for large x and t.

$$K(x,t) = \int_0^\infty A(k,t) \cdot \exp(i(k \cdot x - \omega \cdot t))\, dk \qquad (7)$$

This integral can be put into the form

$$K(x,t) = \int A(\alpha, \beta) \cdot \exp(i \cdot \alpha \cdot \varphi(\beta))\, d\beta \qquad (8)$$

where $\alpha = (\frac{mc^2}{\hbar})t$ and $\beta = \frac{\hbar k}{mc^2}$.

In this case, we can determine each path using the method of stationary phase developed by Kelvin. The basic idea in Kelvin's approach is that the main contribution to the integral comes from the small intervals centered about those values of β for which the phases of the exponential are stationary. Because α is large we have approximately

$$K(x,t) \approx (t^2 - x^2/c^2)^{\frac{3}{4}} \cdot \exp\left(-i\frac{mc^2}{\hbar}\sqrt{t^2 - x^2/c^2}\right) \qquad (9)$$

We discuss the character of the motion furnished by (9) in more detail. Indeed for not too great changes in x or t the waves behave very like simple harmonic waves of a certain period and wavelength. Because the period T correspond to change $2\pi$ in the phase, we have

$$\frac{mc^2}{\hbar}\sqrt{(t+T)^2 - x^2/c^2} - \frac{mc^2}{\hbar}\sqrt{t^2 - x^2/c^2} = 2\pi \qquad (10)$$

Also, we obtain approximate formula for frequency $\omega = \frac{2\pi}{T} \approx \frac{mc^2}{\hbar}\sqrt{t^2 - x^2/c^2}$.

In the same way we find for the local wave length λ=1/k approximate formula

$$\frac{mc^2}{\hbar}\left(\sqrt{t^2 - x^2/c^2}\right) - \frac{mc^2}{\hbar}\left(\sqrt{t^2 - (x+\lambda)^2/c^2}\right) = 2\pi \quad (11)$$

$$\lambda \approx \frac{\hbar\sqrt{1 - x^2/(ct)^2}}{mx/t}$$

As we can see, the position x of group of waves of fixed wavelength at time t is given by the formula

$$x = (\hbar k t/m) \cdot \frac{1}{\sqrt{1 - \frac{\hbar^2 k^2}{m^2 c^2}}} \quad (12)$$

or x=vt, where the group velocity is a constant. In contrast, the phase velocity is not constant. The distinction between these velocities is crucial, and the group velocity plays the dominant role. It determines the propagation of wave vector/frequency and variations in amplitude or wave energy density.

In continuum mechanics, instead of dispersive waves, it is perhaps more precise to speak of a dispersive medium in which the waves are generated. A general approach is usually based on variation principle. The continuity equation is rather the conservation of wave action, which in simple cases becomes energy conservation equation. In those cases where Lagrangian is the difference of kinetic and potential energy, equipartition exists between two stationary values.

The problems of slowly varying wave train in continuum mechanics are analogous to the problems of adiabatic invariant in the mechanics of finite systems. For a wave associated with a free particle the elementary action is ds=h*dθ=L*dt and the energy is E=p*v-L. In this case, the total energy is a constant of integration. For a particle with rest mass m and linear momentum p the total energy is $E^2=p^2c^2+m^2c^4$.

We obtained the propagation formulas from the wave theory that supposed that the wave disturbance caused by deviation in the limited space region is the sum of simple harmonic waves of different amplitudes and wave vectors. Since our medium is a disperse medium the disturbances with increasing time break up into separate trains of waves each of which has approximately the same wave length. Also, we assumed that the probability of finding the particle at a given point and time is proportional to the intensity of associated wave.

This is very close to Feynman's results, which he received another way using the method of path integral. We have in the limit v/c→0 from (9) his amplitude of probability

$$K(x,t) \approx \frac{1}{\sqrt{2\pi k i/m}} \cdot \exp\left(imx^2/2\hbar t\right) \quad (13)$$

Feynman and Hibbs come to this formula from an idea that all paths for particles are possible but not all, however, have the equal probability. For the large particles the contribution of the waves with phases far to stationary phase almost completely cancel one another. Thus, the particle follows a path that satisfies the principle of least action.

The analogy of the material and quantum nonpropagating waves naturally appears when we consider wave disturbances on an infinite strip of width h. If x(z,yt) is the deflection of the strip, the solutions satisfying the certain boundary conditions are x=$e^{iq}$sin($\pi$my/h). These solutions are normal modes with the cutoff frequency $\omega_c$=$\pi$mc/h. Here c is the characteristic speed and m is the order of the mode. A nondispersive mode corresponds m=0.

As we approach wavelengths comparable with interatomic space a, the frequency of the oscillation should be computed in terms of the mass of the atom m and the elastic constant of the strip material βa. To simplify the calculations, we consider it as a one-dimensional string of atoms. Then the force on the nth atom is equal

$$m\ddot{x}_n = \beta(x_{n+1} + x_{n-1} - 2x_n), \quad (14)$$

where $x_n$ is the deviation of the nth atom from the equilibrium position. It can be shown that $x_n$=$A_n e^{i(\omega t+kna)}$ and

$$\omega = \pm\sqrt{\frac{4\beta}{m}} \sin\frac{ka}{2}. \quad (15)$$

We now can find the number of oscillations u(k) with frequencies between k and k+dk. The entire treatment is essentially the same of that used with waves in material strings, electromagnetic fields and nuclear matter. As the length of the string or the lattice constant in the case of the crystal is L, we have in k-space u(k)=L/2$\pi$ and $k_x = \pm 2l\pi/L, k_y = \pm 2m\pi/L, k_z = \pm 2n\pi/L$, where l, m, and n are integers.

Here, the appearance of a discrete set of N-1 allowed values of k in is connected with string structure of N atoms. For wavelengths shorter than the mean distance between the atoms, propagation becomes impossible. The mode density per unit frequency interval is u(k)(dk/dν)dν. We have from (15) that $\frac{dk}{d\nu} = \frac{2}{a\sqrt{\nu_m^2 - \nu^2}}$ and $\nu_m$=(4β/m)$^{1/2}$.

As the string density ρ=m/a, the characteristic speed c=(βa/ρ)$^{1/2}$=a$\nu_m$/2.

We then apply statistical mechanics to these modes to determine the mean energy at the temperature T. The probability that the mode has an energy corresponding to its nth allowed value is $e^{-nh\nu/KT}$ for a given energy $E_n$=nhν. Here n is a positive integer and K is Boltzmann's constant. As the material modes have only one frequency, we obtain the mean energy for the characteristic frequency

$$\overline{E} = h\nu \frac{e^{-h\nu/kT}}{1-e^{-h\nu/kT}}. \quad (16)$$

As a=$10^{-8}$ cm and c=3*$10^5$ cm/s, the maximum frequency $\nu_m$=2c/a of the material oscillation is less than $10^{14}$s$^{-1}$, and the discrete character of energy of the sound waves is usually considered at low temperatures. As we go to higher frequencies, quantization of energy is easier to observe. The assumption that the maximum frequencies of the waves in background medium is about $10^{22}$s$^{-1}$ leads to restriction on its characteristic size a=2c/$\nu_m$=$10^{-16}$ m and energy density $u = 3\frac{\hbar c\pi}{(2\pi)^3} k_m^4 \approx 10^{37}$ N/m$^2$. In an ideal relativistic medium, the pressure p=u/3≈$10^{36}$ N/m$^2$ and longitudinal velocity is c/√3.

## Wave Interpretation of the Maxwell Theory and Newtonian Mechanics

In case of a nonpropagation wave associated with a charged particle moving with the velocity $\vec{v}'$ in the field created by the charged particles resting in the frame K', we must add to the phase or elementary wave action $ds'=\omega_c dt'$ a suplementary term that described this interaction. It must be only scalar from the demand of the phase covariance.

Then the wave action is $ds' = -[mc^2\sqrt{1-v'^2/c^2} + e\phi(\vec{r}',t')]dt'$, where the constant e is characterizing the charged particle and $\phi$ is an arbitrary function. Introducing $dt'$ and $\vec{v}' = d\vec{r}'/dt'$ from (1,2), we obtain in the frame K

$$ds = [-mc^2\sqrt{1-v^2/c^2} + \frac{e}{c}(\vec{A}\vec{v}) - e\phi(\vec{r},t)]dt = L(\vec{r},\vec{v},t)dt, \quad (17)$$

where $\phi(\vec{r},t) = \phi(\vec{r}',t')/\sqrt{1-V^2/c^2}$ and $\vec{A} = \frac{1}{c} \cdot \phi(\vec{r},t) \cdot \vec{V}(\vec{r},t)$.

Now the Maxwell's equations may be derived from the Lagrange equation $\frac{d}{dt}(\partial L/\partial \vec{v}) = \partial L/\partial \vec{r}$, where L in (17) is represented by means of a scalar potential $\phi$ and a vector potential A. We have that the variable conjugates to r is $\frac{\partial L}{\partial \vec{v}} = \vec{p} + \frac{e}{c}\vec{A}$ and the force experienced by the particle is

$$\frac{d\vec{p}}{dt} = -\frac{e}{c}\frac{\partial \vec{A}}{\partial t} - e \cdot grad\phi + \frac{e}{c}[\vec{v} \times curl\vec{A}] \quad (18)$$

It is known that Maxwell perceived his equations as descriptions of the physical characteristics of a certain medium-ether. He described the state of the ether by two directed magnitudes, the electric and magnetic fields, whose changes in space and time are connected by four field equations. The first part of the force in (18), which does not depend on the velocity $\vec{v}$, is called the electric field intensity. The second part is perpendicular to this velocity. It is called the magnetic field intensity.

It can be shown that electromagnetic energy is equal $E=1/8\pi\int(E^2 + H^2)dr$, where integration is over all the space available to the field. To determine the way, in which this energy is distributed among the various frequencies, one can use Fourier analysis. In so doing, one can see that the electromagnetic field behaves like a collection of simple mode $\vec{A}_k = \sum_k [\vec{A}_k(t)\cos\vec{k}\vec{r} + \vec{B}_k(t)\sin\vec{k}\vec{r}]$.

There are an infinite number of field modes, but the infinity is discrete, or countable. The number of possible modes in the element of volume $V=L^3$ in polar coordinates is $dN = \frac{4\pi V}{(2\pi)^3}k^2 dk$. Multiplying the mean energy of oscillators from (16) by dN, we find the Plank distribution of blackbody radiation $u(\nu) = \frac{8\pi V}{c^3}h\nu^3 \frac{e^{-h\nu/kT}}{1-e^{-h\nu/kT}}$.

The electromagnetic fields remain invariant under the gauge transformation
$\vec{A}' = \vec{A} - \nabla \psi$ and $\phi' = \phi + \frac{1}{c}\frac{\partial \psi}{\partial t}$. A common choice is to make divA=0 in empty space.
Then $\vec{k}\vec{A}_k(t) = \vec{k}\vec{B}_k(t) = 0$ and the electromagnetic waves are transverse. Introducing the vector potential and the partial derivative of time from (3), we can rewrite the equations for electrical and magnetic fields

$$\vec{E} = -\frac{1}{c^2}\partial\phi/\partial t \cdot \vec{V} - \frac{1}{c^2} \cdot \phi \cdot \lambda \cdot \vec{V} - grad\phi \quad (19)$$

$$\vec{H} = \frac{1}{c} \cdot \phi \cdot curl\vec{V} + \frac{1}{c}[grad\phi \times \vec{V}] \quad (20)$$

Multiply vector of the electric field intensity by vector velocity, we obtain $[\vec{V} \times \vec{E}] = [grad\phi \times \vec{V}]$. Comparing with equation (20) and taking into account condition that rotor of velocity is zero, we have $\vec{H} = \frac{1}{c}[\vec{V} \times \vec{E}]$. Taking the curl of this formula, we have

$$curl\vec{H} = \frac{1}{c}\{(\vec{E} \cdot grad)\vec{V} - (\vec{V} \cdot grad)\vec{E} + \vec{V} \cdot div\vec{E} - \vec{E} \cdot div\vec{V}\} \quad (21)$$

There are some characteristics that are conserved in continuum mechanics. Take into account condition that the flux of vector of the electric field intensity through the "liquid" surface conserves in time (Helmgoltz's theorem of vortex line consevation $\frac{d\vec{E}}{dt} - (\vec{E} \cdot grad)\vec{V} + \vec{E} \cdot div\vec{E} = 0$), we obtain

$$curl\vec{H} = \frac{1}{c}(\frac{\partial \vec{E}}{\partial t} + \vec{V} \cdot div\vec{E}) \quad (22)$$

Originally, Maxwell called the εμ∂E/∂t the displacement current because the constant ε is associated with volume polarizability of free space. As ε=8.85*10⁻¹² F/m and μ=4π*10⁻⁷H/m, we have c=1/(εμ)^(1/2)=3.00*10⁸m/s. Taking the divergence of both sides of (22), we have $\frac{\partial \rho}{\partial t} + div(\rho \cdot \vec{V}) = 0$, where we introduce new function ρ. It satisfies the law conserve or the continuity equation, where $\vec{V}(\vec{r},t)$ is the group velocity of de Broglie wave connected with the particles created field. Then ρ is the charge density that equal the charge multiplying on the probability density of finding the particles at a given point and time.

Another condition is that the integral from scalar potential by "liquid" volume conserves in time (Lorentz condition of classical electrodynamics $\frac{\partial \phi}{\partial t} + div\vec{A} = 0$). Taking the divergence of vector $\vec{E}$ and taking into account the Lorentz condition we can derive the wave equations for the scalar and vector potentials.

If the transverse waves correspond to the electromagnetic field, it seems possible that longitudinal waves correspond to the gravitational field. By adding to the elementary action a supplementary term $-m\phi(\vec{r}',t')$, we can compose the Lagrange function of wave associated with a particle of mass m moving with the velocity $\vec{v}'$ in the wave field created by the particles resting in the frame K'.

In the time-independent and irrotational case ($\partial A/\partial t$ and curl($\vec{A}$)=0), by analogy with the equations (17, 18) we have the gravitational field intensity E=grad$\phi$. For the large positive mass M, $\phi$ is the Newton potential -GM/r, where the appearance of a constant G is a choice of units. Consider a system with one degree of freedom we can find the law of planet orbital motion from the equation (6) in polar coordinates

$$-\frac{1}{c^2}(\frac{\partial S}{\partial t}-\frac{GM}{r})^2 + (\frac{\partial S}{\partial r})^2 + \frac{1}{r^2}(\frac{\partial S}{\partial \varphi})^2 + m^2c^2 = 0. \quad (23)$$

From (23) follows that only a small variation from the classical law of motion of a planet takes place in this approach $\delta\psi = \frac{2\pi^2 GM}{c^2 p(1-e^2)}$, where $\delta\psi$ is displacement of the perihelion of planetary orbits per revolution, p is the semi-major axis, e is the eccentricity. It is convenient to think of relativistic medium energy in terms of mass density. From dimensional analysis one might expect that the characteristic wave speed is $c^2$=p/$\rho$. The assumption that it is equal to the velocity of light leads to restriction on equivalent mass density $\rho$=p/$c^2$≈$10^{17}$ kg/m$^3$.

**Strong Forces and Cosmic Ray Acceleration**

From a classical point of view, it is necessary to introduce forces that hold the Coulomb repulsive forces of the electrical charge in equilibrium. At first, Poincare introduced a scalar cohesive pressure that holds electrons together. A classical view of the electron pictures it as a sphere, whose rest mass energy mc$^2$ is the energy stored in its electrical field. For the relativistic medium pressure of about $10^{36}$ N/m$^2$, the electron radius is about 0.1fm ($10^{-15}$ m).

Recently, Aspden calculated the proton mass. He suggested that the background medium might have a simple cubic lattice of negatively charged particles. These particles describe a circle in a background continuum of positive charges. Aspden received that energy quantum Q and proton mass M are connected by formula M=(4+2*$6^{1/2}$)*Q. Here the quantum Q is equal to the energy contained within a lattice cell and is slightly smaller than the mass energy of muon. It is associated with the cohesive pressure required to keep lattice particles from disintegrating [3].

Following Jones, we assumed that the pressure of relativistic medium produced the strong but short-range attractive force. For example, a 50,000-N strong force holds quarks together to form protons. As the particles separate and the medium follows between them, the binding force goes to zero. A 1,000-N residual effect of medium pressure holds the protons and unstable neutrons together to form the nucleus. Balance of binding nuclear and repulsive electromagnetic forces determines the structure and characteristics of the stable nuclei.

The electric force becomes dominant at separation of more than a few fm. We can describe this effect in terms of the potential energy. To a first approximation for simple nuclei, it may be represented by the infinite square well. Taking into account the Pauli exclusion principle, we have for a nucleus with the A nucleons and the total number of the modes $k^2 \leq K^2_F$ that $A = \frac{4V}{(2\pi)^3}\frac{4\pi}{3}K^3_F$.

All nuclei have essentially the same density ρ and binding energy $E_b$. As their radii are approximately $R=R_0 A^{1/3}$, then the density of nuclear matter is $\rho = A/V = 3/4\pi R_0^3$ and $K_F = 1.52/R_0$. As $R_0 = 1.2$ fm and $E_b = 8$ MeV, we have that the nuclear well potential energy $U = E_F + E_b = 42$ MeV.

The relativistic dispersion relation or Dirac's formula for particle energy leads to the existence of quantum states with negative energy and negative-mass particles (modes with negative group velocity). According the Newton's laws antiparticles, which have the direction of their acceleration opposite to that of the force, are accelerated toward the positive mass. But they repulse each other gravitationally. Therefore, antiparticles tend to fill the entire cosmic space with approximately homogenous density.

In this wave model, the fundamental distinction in matter and antimatter gravitational acceleration could be responsible for the present asymmetric situation. The universe began as a giant sphere, which contained equal amount of matter and antimatter (the total internal energy is zero). Enormous amount of energy and radiation resulted from their interactions. When galaxies began to condense and the net gravitational attraction slightly reduced, the universe started expanding. The expansion cooled not only matter but also radiation.

Obviously stars cannot contain a close mixture of matter and antimatter. However, no observations presently exclude the presence of the large amount of antimatter concentrated in the outer part of the universe or trapped between the widely separated clusters of galaxies. For the galaxies the gravitational force acting between them could consist of the ordinary attractive force and repulsive force associated with antimatter expansion.

We consider an arbitrary spherical shell of matter expanded with antimatter. If the shell with radius R encloses a mass M, gravitational potential energy of a test particle (for example, galaxy) is -GM/R per unit mass. The kinetic energy of expansion is $v^2/2$ per unit mass. The recession speed v could be expressed with the Hubble law as v=HR, where H is the constant. From the conservation of energy principle we have

$$\frac{H^2}{2} - \frac{4\pi G \rho}{3} = \frac{u}{R^2} \quad , \qquad (24)$$

where u is the internal energy of the universe per unit mass.

As electric fields are rare in the fully ionized interstellar plasma, cosmic ray energy conserves in the cosmic space. The antimatter cannot be a closed system in thermodynamic equilibrium. Some long-range interaction between the two types of matter must present. In this case, observed energy spectrum of the primary cosmic rays might be partially explained by its interactions with matter/antimatter.

To develop this point of view, we consider a moving particle of kinetic energy E, charge ze and velocity v. On the interactions with the distant electrons and positrons both a loss and gain of energy might take place. This energy gain must come from the negative kinetic energy of positrons, when their energy is reduced. Gain of energy is larger than losses because the traditional particles are more condensed.

By an interaction with the positrons at distance q the particle of the charge ze will gain of energy

$$\frac{dE}{dx} = k\frac{4\pi z^2 e^4 n}{mv^2}\ln\frac{q_{max}}{q_{min}}, \qquad (25)$$

Here the factor k is introduced to account for the fact that matter is more condensed than antimatter, $q_{min/max}$ are the distance of closest and more distant approach, n is the number of electrons/positrons per unit volume.

We assume that the closest distance is equal $e^2/E$ cm. The maximum distance is equal the Debye length $\lambda = \sqrt{\frac{KT}{4\pi n e^2}}$, beyond which the Coulomb field of the fast particle is screened by collective motion of the electrons/positrons. If the coefficient k=0.5, the density of electron-positron pairs with $KT/mc^2=0.01$ is n=1cm$^{-3}$, the rate of energy gain is about $10^{-17}$L ev/cm for protons and $5*10^{-15}$L ev/cm for iron nuclei, where L is the size of the acceleration region.

The mean energy acquired by a particle is a product of an absorption mean free path L, charge ze and the gain of energy per unit path E=zeLdE/dx. The cosmic proton energy gain in time is a product of dE/dx, their density $n_c$ and the velocity of light c, then dW/dt=cn$_c$dE/dx. The energy density gain occurs whenever the relativistic particle concentration exists relatively close to antimatter. For example, for a space region of volume $10^{45}$cm$^3$ with $n_c=10^{-3}$ cm$^{-3}$ the energy output could be about $10^{34}$ W [4].

If the place of the production and acceleration of the cosmic rays up to energy $10^{15}$ev is a supernova (SN) explosion in matter-rich part of the universe, the particles escaping into outer antimatter-rich part of the universe can gain kinetic energy and then pass back with energy up to $10^{18}$ev. A cloud of antimatter extending 3,000 light years above the center of our Galaxy may be the source of cosmic ray acceleration to energy above the GZK cutoff ($10^{20}$ev). In this approach, iron nuclei could dominate in the ultra high-energy cosmic ray flux.

Since the galactic SN rate is two per century and cosmic ray lifetime is $2*10^7$ years, it should be a stationary cosmic ray flux. For a sphere with uniform density of sources S and macroscopic absorption cross-section $\Sigma_a$, the average flux is F≈S/π*($\Sigma_a$+DB$^2$) and leakage probability P=B$^2$L$^2$/(1+ B$^2$L$^2$). Here the buckling B=π/R and the diffusion mean free path L$^2$=D/$\Sigma_a$. If D=$2*10^{18}$cm, R=$10^{30}$cm, $\Sigma_a=2*10^{-25}$cm$^{-1}$, L=$10^{22}$cm and S≈$10^{-24}$ p/s*cm$^3$, we have average cosmic ray flux F≈1.5 p/s*cm$^2$ and leakage probability P≈$10^{-15}$.

This approach might be analyzed with the recent observations of the gamma-ray burster large energy output, the ultra high-energy cosmic ray anisotropy and the Fermi's interpretation of the muon-decay experiments in mind. In the muon-decay experiments, the absorption of muons that are created by cosmic rays in Earth's atmosphere is larger than that of equal masses of condensed material. Fermi suggested that screening properties of the medium are an important factor in the interpretation observed difference in muon absorption in air and condensed materials [5].

**Radio Astronomical Analysis of the Velocity Effects on Atomic Clocks**

The traditional relativistic experiments are analyzed here as special cases of the more general effect-Sagnac effect on open paths. The uniform translation is simply a limited case with infinitely great radius, and the earth's rotation around the sun was often treated as a translation in the relativistic experiments. On another hand, with two signals starting at one point, we have the traditional Sagnac effect on closed path.

In the Sagnac experiments, the two signals are sending in the opposite directions along the same path S=φR, where φ is central angle and R is radius. When the system is set in rotation with angular rate Ω, path length differences will result for the two signals $\Delta S_{1,2} = \Omega R t_{1,2} = v t_{1,2}$ (see Fig.1).

It usually had been considered that Sagnac measured only anisotropy angular velocity not linear. As $t_1 = (S+\Delta S_1)/c$, $t_2 = (S-\Delta S_2)/c$ we have

$$\Delta t = \frac{S}{c-v} - \frac{S}{c+v} = \frac{2 \cdot S \cdot v}{c^2} = \frac{4 \cdot A \cdot \Omega}{c^2}, \qquad (26)$$

where $A = \varphi \cdot R^2 / 2$ is area of the circular sector.

In the case of the Sagnac effect on open paths connected with atomic clock synchronization, the two effects are measured simultaneously: difference of clock frequencies and difference of transmission times. The atomic clock is based on condition $\Delta E = \hbar \nu$ and essentially represents a frequency standard. A wave train results when a periodical signal lasts for a finite time. To simplify we consider that the train propagates in x direction and consists of exactly n waves of wavelength λ=c/ν.

If the train reaches the point $x_0$ at the $t_0$ and leaves $x_1$ at the time $t_1$, we have $c(t_1-t_0)=x_1-x_0+n\lambda$ and $n=\nu(t_1-t_0 -\Delta x/c)$. By using Galilean transformation we can compute n exactly in the same way in a moving system

$$n = \nu'(t_1 - t_0 - \frac{x_1' - x_0'}{c'}) = \nu(t_1 - t_0 - \frac{x_1 - x_0}{c}) \qquad (27)$$

We have from (27) that frequency and velocity change proportionally due to motion of light receiving systems ν/c=ν'/c'. The GPS and radio astronomy techniques might provide an independent experimental procedure to estimate the frequency effect. For measuring the frequency effect of the moving atomic clocks, we consider that the atomic clocks are associated with the radio telescopes. The simple digital electronics can be used at the experiments with ground or satellite clocks (Fig.2).

The received signals come to the counter controlled by the scaler. The scaler counts local clock pulses and after a certain predetermined number has been counted, it sends a short start/stop signal to the counters and data acquisition system. To test the electronics setup stability as a function of time of day several 4-hours runs using cable connection were performed. To define whether the period of the clock's frequency variations is the solar or sideral day the measurements extending over several hours will be made periodically during the day. Two methods might be used: observations one source close to the ecliptic pole and observations of pair sources of equal elevation [6].

The Sagnac effect holds for any type of waves. It can be shown that the Sagnac phase difference for matter waves is $4Am\Omega/h$. Since $\upsilon = mc^2/h$, it agrees with the formula for the optical Sagnac effect. However, the matter-wave interferometer is more sensitive to rotation by the factor $c/v$, and in the non-relativistic limit, the Sagnac phase shift of matter waves is independent of the wavelength.

Since rate of a clock in the gravitational field changes, the clock located in the balloons/satellite could experience the shift relative to a clock located on the earth. Singer originally suggested the digital method to test the general theory of relativity by measuring gravitational red shift of the satellite clocks [7].

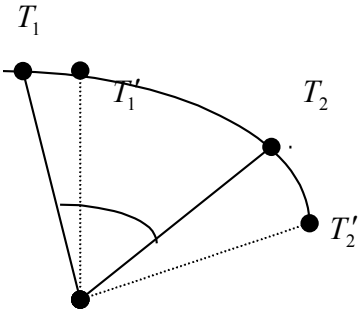
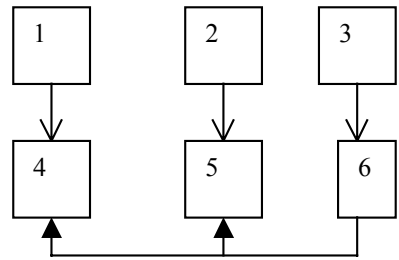

**FIGURE 1.** Simplified configuration of the Sagnac experiment.

**FIGURE 2.** Electronics setup. 1,2-Receivers; 3-Atomic clock; 4,5-Counters; 6-Scaler.

## Conclusion

We came to the highly probable conclusion that fundamental fields are the wave disturbances in certain dispersive medium that defines the preferred reference frame. Originally, the quantum theory and relativity began with the study of blackbody radiation and anisotropy of velocity of electromagnetic waves. When the quantum hypothesis leads to excellent agreement with the experiment, the most debated problem in space-time physics is disparity between the Michelson and Sagnac experiments. Also, it is difficult to dismiss the concept of absolute space, where there is inertia.